# A Dynamic Clustering-Based Markov Model for Web Usage Mining

José Borges
School of Engineering, University of Porto, Portugal, `jlborges@fe.up.pt`

Mark Levene
Birkbeck, University of London, U.K., `mark@dcs.bbk.ac.uk`

**Abstract**

Markov models have been widely utilized for modelling user web navigation behaviour. In this work we propose a dynamic clustering-based method to increase a Markov model's accuracy in representing a collection of user web navigation sessions. The method makes use of the state cloning concept to duplicate states in a way that separates in-links whose corresponding second-order probabilities diverge. In addition, the new method incorporates a clustering technique which determines an efficient way to assign in-links with similar second-order probabilities to the same clone. We report on experiments conducted with both real and random data and we provide a comparison with the $N$-gram Markov concept. The results show that the number of additional states induced by the dynamic clustering method can be controlled through a threshold parameter, and suggest that the method's performance is linear time in the size of the model.

## 1 Introduction

Modelling user web navigation data is a challenging task that is continuing to gain importance as the size of the web and its user-base increase. Understanding the way users navigate (or surf) the web helps in formulating guidelines for web site reorganization, in facilitating the combination of caching with prefetching of web pages in order to reduce web latencies, in providing the user data needed to build adaptive web sites, in making available statistics for effective placement of web advertising, in providing browsing information from users for enhancing search engines, and in providing the data that forms the basis for understanding and influencing web users' buying patterns [Srivastava et al., 2000, Mobasher, 2004].

One way to characterize past user web navigation is through information collected from server log files. A log file provides a list of page requests made to a given web server, where a request is characterized by, at least, the IP address of the machine placing the request, the date and time of the request, and the URL of the page requested. From this information it is possible to reconstruct the user navigation sessions within the web site [Berent et al., 2001, Spiliopoulou et al., 2003], where a session consists of a sequence of web pages viewed by a user within a given time window. The web site owner can take advantage of web usage mining techniques to gain insight about users' behavior when visiting the site and to use the acquired knowledge to improve the design of the site.

Several authors have proposed the use of Markov models to represent a collection of user navigation sessions on the web. Pitkow et al. [Pitkow and Pirolli, 1999] proposed a method to induce the collection of longest repeating sub-sequences from log data. Assuming that the longest sub-sequences contain more predictive power, a Markov model is then inferred from such sequences. Sarukkai [Sarukkai, 2000] presents a study showing that Markov models have prediction power and on this basis proposes a system based on such a model for predicting the next page accessed by the user. Cadez et al. [Cadez et al., 2000] utilize Markov models for classifying browsing sessions into different categories. A model-based clustering approach is used in which users with similar navigation patterns are grouped into the same cluster (each cluster is represented by a Markov model), and a method to visualize the data on each cluster is also presented. Deshpande et al. [Deshpande and Karypis, 2001] propose techniques for combining different order Markov models to obtain low state complexity and improved accuracy. The method starts by building the $All - K^{th} - Order$ Markov model, that uses the highest order model out of $K$ that covers each state, and makes use of three different techniques to eliminate states in order to reduce the model complexity. Zhu et al. [Zhu et al., 2002] propose to use a Markov model inferred from user navigation data to measure page co-citation and coupling similarity, based on in-link and out-link similarity. A clustering algorithm is then used to construct a conceptual hierarchy and the Markov model is used to estimate the probability of visiting another cluster of pages in the concept hierarchy. Jerpersen et al. [Jespersen et al., 2003] study the quality of a fixed-order Markov model in representing a collection of navigation sessions and, according to two proposed measures for the quality of a set of patterns, they conclude that a fixed-order Markov model has some limitations in the accuracy achieved.

An alternative approach to model user web navigation sessions is the use of tree-based models. Schechter et al. [Schechter et al., 1998] use a suffix tree that represents the collection of paths inferred from log data to predict the next page accessed. Dongshan and Junyi [Dongshan and Junyi, 2002] propose a hybrid-order tree-like Markov model to predict web page access, which provides good scalability and high coverage. The model is built using several tree-like Markov models, where each tree corresponds to a given Markov order and the prediction is determined by a voting scheme. Chen and Zhang [Chen and Zhang, 2003] propose to build a PPM (Prediction by Partial Match) tree that only allows popular nodes to be root nodes. Based on the assumption that most user sessions start with popular pages they reduce the complexity of the model (measured in the number of nodes in the tree) by eliminating branches that have non-popular pages as their root. Finally, Gündüz and Özsu [Gündüz and Özsu, 2003] propose to cluster user sessions based on a given similarity measure. The user sessions are clustered to reduce the search space of a recommendation engine that, for each cluster, makes use of a click-stream tree to which the new user session is matched.

Markov models have been shown to be suitable to model a collection of navigation records, where higher order models present increased accuracy but with a much larger number of states. In previous work we proposed to model a collection of user web navigation sessions as a Hypertext Probabilistic Grammar (HPG) [Borges, 2000, Borges and Levene, 2000]. A HPG corresponds to a first-order Markov model which makes use of the $N$-gram concept, [Charniak, 1996], in order to achieve increased accuracy; for the full details on the HPG concept see [Borges and Levene, 2000]. In [Borges and Levene, 2000] an algorithm to extract the most frequent traversed paths from user data was proposed, and in [Borges and Levene, 2004] we have shown that the algorithm's complexity is, on average, linear time in the number of states of the grammar.

Here we propose a novel method to increase a Markov model's accuracy in representing a collection of user web navigation sessions. The proposed model makes use of the state cloning concept in order to accurately represent second-order transition probabilities and in some cases higher order probabilities. The concept of state cloning in the context of usage mining was first introduced in [Levene and Loizou, 2003], where a method is proposed that applies the cloning operation to states whose first and second-order probabilities diverge, by duplicating such states in a way that separates their in-links. As defined, the method given in [Levene and Loizou, 2003] aims at accurately representing second-order probabilities, which take into account the history of the web page a user came from prior to clicking on a link leading to another page. With such method, what changes is the probability of arriving at a given state and not the probabilities of leaving that same state. Here we propose a new cloning condition, a new definition for the cloning operation, and a different interpretation for second-order probabilities. Our cloning method incorporates a clustering technique to determine an efficient way to assign in-links with similar second-order probabilities to the same clone. An extension of this concept to higher order probabilities will be presented in a subsequent publication.

In Section 2 we present the background on which the method we propose is based. Section 3 presents our new dynamic clustering-based method and Section 4 presents results of experiments conducted with both synthetic and real data. In Section 5, we present preliminary evaluation of the method, and finally, in Section 6 we give our concluding remarks.

## 2 Background

### 2.1 Building the Model from Navigation Sessions

In previous work we have proposed to model a collection of user web navigation sessions as a HPG [Borges and Levene, 2000, Borges, 2000]. A HPG is a self-contained and compact model which is based on the well established theory of probabilistic grammars providing a sound foundation for enhancements, including a detailed study of its statistical properties. A HPG corresponds to a first-order Markov model inferred from a collection of user web navigation sessions, which we now review with the aid of an example.

Consider a web site composed by six pages, $\{A_1, A_2, \ldots, A_6\}$, and the collection of navigation sessions given in Table 1 (NOS represents the number of occurrences of each session).

A navigation session gives rise to a sequence of pages viewed by a user within a given time window. To each web page visited there corresponds a state in the HPG model. Moreover, there are two additional states: a start state, $S$, representing the first state of every navigation session; and a final state, $F$, representing the last state of every navigation session. There is a transition corresponding to each pair of pages visited in sequence in a session, a transition from the start state $S$ to the first state of a session, and a transition from the last state of the session to the final state $F$. The model is incrementally built by processing the complete collection of navigation sessions.

| Session | NOS |
|---|---|
| $A_1, A_2, A_3$ | 3 |
| $A_1, A_2, A_4$ | 1 |
| $A_5, A_2, A_4$ | 3 |
| $A_5, A_2, A_6$ | 1 |

Table 1: A collection of user navigation sessions.

The probability of a transition is estimated by the ratio of the number of times the corresponding sequence of states was traversed and the number of times the anchor state was visited. In order to set the initial probabilities we make use of a parameter $\alpha$ by which we can tune the model to be between a scenario where the initial probabilities are proportional to the number of times a page has been requested as the first page and a scenario where the probabilities are proportional to the number of times the page has been requested. If $\alpha = 0$ only states which were the first in a session have probability greater than zero of being in a transition from the start state, on the other hand if $\alpha = 1$ the initial probability of a state is estimated as the proportion of times the corresponding page was requested by the user. Therefore, when $\alpha > 0$, according to the model's definition, apart from the final state, all states have a positive initial probability.

Figure 1 represents the HPG model corresponding to the collection of sessions given in Table 1 and for $\alpha = 0$. Next to a transition, the first number indicates the number of times the corresponding hypertext link was traversed and the number in parentheses represents the estimated transition probability.

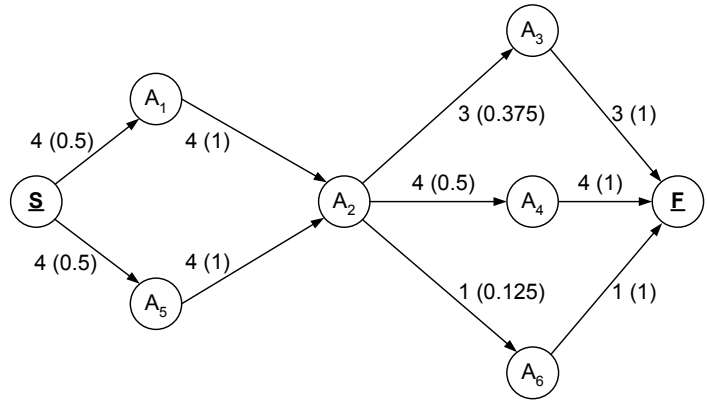

Figure 1: The first-order model corresponding to the sessions given in Table 1.

Given a model with the set of states $\{S, A_1, ...., A_n, F\}$ we let $w_i$ represent the number of times the page corresponding to state $A_i$ was traversed, $w_{i,j}$ be the number of times the hypertext link from page $A_i$ to page $A_j$ was traversed, $w_{S,i}$ be the number of times page $A_i$ was the first in a navigation session and $w_{i,F}$ be the number of times a session terminated in page $A_i$. In addition, we let $p_{i,j}$ be the first-order transition probability from state $A_i$ to $A_j$, $p_{S,i}$ be the first-order probability of the transition from the starting state $S$ to state $A_i$ and $p_{i,F}$ be the first-order transition probability from state $A_i$ to the final state $F$. In Definition 1 we give the formal definition for the first-order transition probabilities.

**Definition 1 (first-order transition probability)** *Given a HPG having states* $\{S, A_1, \ldots, A_n, F\}$ *the first-order transition probabilities are estimated according to the following expressions:*

*(i)* $p_{S,i} = \alpha \frac{w_i}{\sum_{j=1}^{n} w_j} + (1 - \alpha) \frac{w_{S,i}}{\sum_{j=1}^{n} w_{S,j}}$,

*(ii)* $p_{i,j} = \frac{w_{i,j}}{w_i}$,

*(iii)* $p_{i,F} = \frac{w_{i,F}}{w_i}$.

As described, the model assumes that the probability of a hypertext link being chosen depends solely on the contents of the page being viewed. Several authors have shown that models that make use of such an assumption are able to provide good accuracy when used to predict the next link the user will choose to follow; see for example [Sarukkai, 2000]. However, in the Section 2.2 we review the $N$-gram concept as proposed in [Borges and Levene, 2000], which was shown to increase the model's accuracy and in Section 3 we propose a new model which relaxes the first-order Markov assumption.

## 2.2 $N$-gram Model

In [Borges and Levene, 2000] we proposed the use of the $N$-gram concept, [Charniak, 1996], to improve a HPG accuracy in representing a collection of user sessions. The $N$-gram concept enables us to fix the user's memory when navigating the web by assuming that only the previous $N-1$ pages visited affect the probability of the choice of which page to visit next. In other words, it assumes that the choice of the page to browse next does not depend on all the history of pages previously seen, but only on the last $N-1$.

From a collection of user navigation sessions the relative frequency with which each page was requested determines its zero-order probability, that is, the probability of a page being chosen independently of any previously visited pages. A 1-gram model is inferred by computing such probabilities for every page. Similarly, the first-order probabilities are estimated by computing the relative frequencies of all the 2-grams in the collection of sessions. A 2-gram is no more than a pair of consecutive requests. With such model, the probability of the next choice depends solely on the current position and is given by the frequency of the 2-gram divided by the overall frequency of all 2-grams with the same initial position. In addition, the initial probability of a state, $A_x$, is proportional to the frequency of the 2-gram $SA_x$ and the probability of a transition from state $A_y$ to state $F$ is proportional to the frequency of the 2-gram $A_yF$. A 2-gram model corresponds to the first-order Markov model. A second-order model is inferred by computing the relative frequencies of all 3-grams and higher orders can be computed in a similar way. Figure 2 shows the 3-gram HPG model corresponding to the collection of sessions given in Table 1.

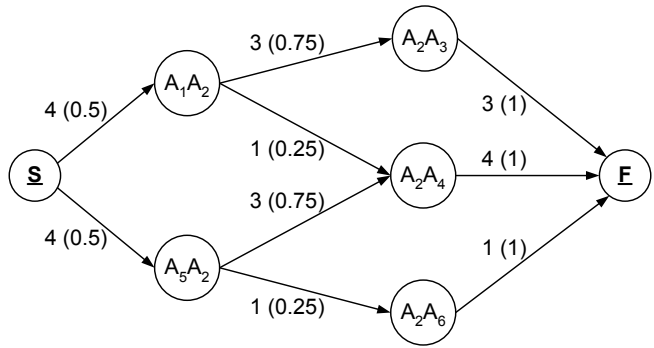

Figure 2: The HPG 3-gram model corresponding to the sessions given in Table 1.

In a $N$-gram model each state corresponds to a navigation sequence composed of $N-1$ pages. Therefore, a $N$-grammar does not generate strings, $w$, of length shorter than the history depth, $|w| < N-1$; similarly, only navigation sessions whose length is greater or equal than $N-1$ can be used in the grammar's construction. For example, when building a 3-grammar all navigation sessions of length 1 are dropped. (Note that a session of length 2 induces the 3-grams composed by the session plus the start state $S$ and also the session plus the final state $F$.)

The drawbacks of the $N$-gram model are the exponential increase in the number of states as the history depth increases and the necessity to assume the same history depth for every state. In [Borges, 2000] we have conducted a study with real data whose results suggest that, in most cases, a second order model provides a good trade-off between the model accuracy and its size. In addition, in Section 4 we study, through a set of experiments conducted with both real and synthetic data, the effect of varying the number of states while increasing the model's order.

## 3  A Dynamic Clustering-Based Markov Model

Here we present a novel method that makes use of the cloning operation concept in order to increase the accuracy of a HPG in representing a collection of user web navigation sessions. The work is inspired by the cloning method described in [Levene and Loizou, 2003] but we propose a different cloning condition, a different way to compute the weight of an out-link for the cloned states and we utilize a clustering technique as an efficient way to clone a given state. The aim of our method is to achieve improved accuracy within the HPG model while controlling the number of additional states.

According to the method we are proposing, a state needs to be cloned if it is not *accurate*, that is, if its first and second-order probabilities diverge. We now present our definition of second-order probability.

**Definition 2 (second-order transition probability)** *We let $p_{i,k\,j}$ be the second-order transition probability, that is, the probability of the transition $(A_k, A_j)$ given that the previous transition that occurred was $(A_i, A_k)$. The second-order probabilities are estimated as follows:*

$$p_{i,k\,j} = \frac{w_{i,k,j}}{w_{i,k}} ,$$

where $w_{i,k,j}$ and $w_{k,j}$ stand for the corresponding 3-gram and 2-gram counts.

Our definition for state accuracy now follows.

**Definition 3 (state accuracy)** *Given a state $A_x$, having $I$ in-links and $O$ out-links, and an accuracy threshold parameter, $\gamma$, with $0 \leq \gamma \leq 1$, if for any $i, 1 \leq i \leq I$, and any $o, 1 \leq o \leq O$, we have that $|p_{i,x\,o} - p_{x,o}| < \gamma$ the state is said to be accurate.*

In addition, we present our definition for model accuracy.

**Definition 4 (model accuracy)** *We say that a model is accurate if every state is accurate.*

In the example shown in Figure 1 the first-order probability of transition $(A_2, A_3)$ is $p_{2,3} = 0.375$. However, from the collection of user sessions (see Table 1), if a user is in state $A_2$ after previously visiting state $A_1$ the probability of following to state $A_3$ is $p_{1,2\,3} = 0.75$. According to Definition 3, and for any $\gamma < 0.375$, state $A_2$ is not accurate and, therefore, the model is not accurate. The application of the cloning operation to state $A_2$ enables us to increase the model accuracy and, when $A_2$ is cloned, state $A_2'$ is created. Since state $A_2$ has two in-links each of the in-links is assigned to one of the states.

We now describe the method we propose to distribute the transition counts among a set of state clones.

**Definition 5 (state cloning)** *We let $A_x$ be a state with $I$ in-links and $O$ out-links that is cloned based on link $(A_1, A_x)$, and $A_x'$ be its clone. The HPG should be modified as follows:*

1. *for every out-link $(A_x, A_o)$, with $1 \leq o \leq O$, add the link $(A_x', A_o)$,*

2. *remove in-link $(A_1, A_x)$ and add in-link $(A_1, A_x')$,*

3. *for every 3-gram $A_1 A_x A_o$ with $1 \leq o \leq O$ do*

    *(a) $w_{x,o} = w_{x,o} - w_{1,x,o}$,*

    *(b) $w_{x',o} = w_{x',o} + w_{1,x,o}$.*

For the HPG given in Figure 1 when state $A_2$ is cloned on the basis of link $(A_1, A_2)$ the result is the HPG given in Figure 3. The transition counts are updated as follows: $w_{2',3} = w_{1,2,3} = 3$, $w_{2',4} = w_{1,2,4} = 1$, $w_{2,4} = w_{2,4} - w_{1,2,4} = 3$, $w_{2,3} = w_{2,3} - w_{1,2,3} = 0$ and $w_{2,6} = w_{2,6} - w_{1,2,6} = 1$.

Note that the resulting model is such that the second-order probabilities are represented exactly as given by the user navigation data. However, if the state to clone has more than two in-links we need a clever way to distribute the in-links between the two states (the original state and its clone) in order to achieve good accuracy. In fact, if a state has three or more in-links and the cloning method is applied repeatedly its is possible to end-up with as many clones of that state as the number of initial in-links minus one. Although the resulting second-order probabilities may be exact it may be possible to achieve a better tradeoff between accuracy and the number of additional states.

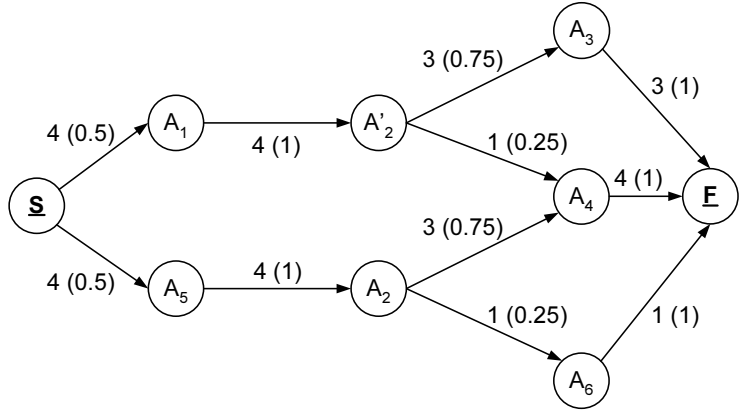

Figure 3: The model resulting from cloning state $A_2$ in Figure 1; in this case the 2nd order probabilities are accurate.

The example given in Figure 4 makes evident the necessity to provide a clever method to distribute the in-links between a state and its clone. Figure 4 (a) represents the first-order model corresponding to the given collection of sessions (states $S$ and $F$ are omitted for the sake of simplicity), Figure 4 (b) represents the model after state $A_2$ is cloned on the basis of link $(A_1, A_2)$ and Figure 4 (c) after state $A_2$ is cloned on the basis of link $(A_5, A_2)$. Note that in Figure 4 (c) the weight of out-links from state $A'_2$ are identical to the weight of out-links from $A''_2$. Therefore, it is possible to obtain identical accuracy with a smaller number of states.

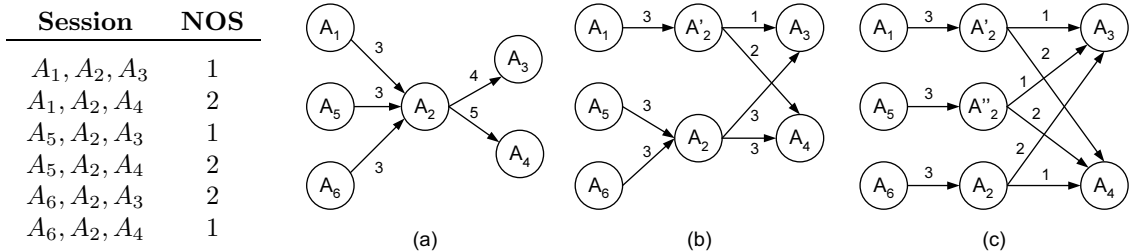

Figure 4: An example to justify the use of a clustering technique in the cloning operation.

States representing a given page and showing identical out-link transition probabilities can be easily identified in order to determine that such states should be merged. However, in the general case, it would be interesting to provide a flexible condition allowing the merging of clones with similar transitions probabilities, and to obtain a good compromise between the model's number of states and its accuracy. We propose to incorporate clustering into the cloning operation in order to provide a method that, for a given state, creates a number close to the minimum of clones necessary to achieve a user specified second-order probability accuracy.

First we will state a set of conditions that need to be met in a state in order to be eligible for cloning.

**Definition 6 (conditions for cloning)** *A state $A_x$ with $I$ in-links and $O$ out-links (see Figure 5), is eligible for cloning if all of the following conditions hold:*

1. *$O > 1$, that is, the state has at least two out-links otherwise the transition probability from it will always be equal to 1;*

2. *$I > 1$, that is, the state has at least two in-links;*

3. *$w_x > V$, where $V$ is a parameter stating the number of times a state has to be visited for the estimation of its probabilities be considered reliable. (In general, we set $V \geq 30$ due to the law of large numbers);*

4. *There is at least one transition $(A_j, A_x)$, with $1 \leq j \leq I$, and one transition $(A_x, A_k)$, with $1 \leq k \leq O$, such that $|p_{j,x\,k} - p_{x,k}| \geq \gamma$, that is, the state is not accurate according to Definition 3.*

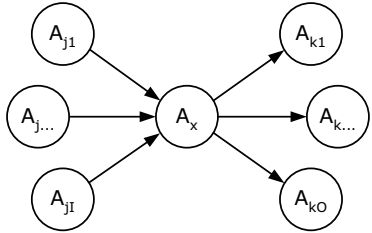

Figure 5: An example of a state $A_x$ having $I$ in-links and $O$ out-links.

We now present our clustering-based cloning method. Consider a first-order HPG with $N$ states that was inferred from a collection of navigation sessions, where we keep the 3-gram counts obtained from the user data in an hash-table. For each state $A_x$, with $1 \leq x \leq N$, assess the conditions for cloning according to Definition 6.

In order to assess the fourth condition, the first-order probability, given by the corresponding transition probability, is compared to the second-order probability inferred from the 3-grams. In more detail, for a state $A_x$, as for example in Figure 5, traverse its $I$ in-links, one at a time, and for each in-link infer all the 3-grams induced by the model's topology. Then, get each induced 3-gram's count from the hash-table and compute the estimate for the corresponding second-order probability. Finally, compare the second-order transition probability to the corresponding first-order transition probability. We note that, transitions to state $F$ and from state $S$ are also evaluated. For example, in Figure 5 and for the in-link $(A_{j_1}, A_x)$ the 3-grams $A_{j_1} A_x A_{k_1}, \ldots, A_{j_1} A_x A_{k_O}$ are induced. Moreover, $p_{j_1,x\,k_1} = w_{j_1,x,k_1}/w_{j_1,x}$. The clustering-based cloning method of state $A_x$ in Figure 5 is carried out as described in Figure 6.

In Figure 7 we give another example of a collection of user navigation sessions and the corresponding first-order HPG model. The clustering-based cloning method is illustrated in Figure 8. In Figure 8 (a), the vector next to state $A_5$ represents the first-order probabilities of transitions from state $A_5$, vectors next to states $A_1$, $A_2$, $A_3$ and $A_4$ represent the corresponding second-order transition probabilities. The evaluation of the link $(A_1, A_5)$ determines that state $A_5$ needs to be cloned based on link $(A_1, A_5)$ as long as the accuracy threshold is $\gamma < 0.67 - 0.5$. Figure 8 (b) shows the first iteration of the method where link $(A_1, A_5)$

1. For each in-link to $A_x$ compute all the second-order probabilities and store them in a vector;

2. Apply the $K$-means algorithm to identify clusters of such vectors:

   (a) Start with $K = 2$ ($K$ indicates the number of clusters), where each cluster corresponds to a state;

   (b) Uniformly at random, assign each in-link to a cluster with the only restriction of making sure that identical vectors are placed in the same cluster;

   (c) Compute clusters centroids, where a centroid is defined by the transition probabilities resulting from the distribution of the weights of the in-links assigned to the cluster according to Definition 5;

   (d) For each in-link, if the closest centroid represents another cluster move the in-link to that cluster. Recompute the new centroids and repeat the process until no further in-link moves occur, meaning that a solution was obtained;

   (e) Assess if the obtained solution is accurate. A solution is said accurate if, for every cluster, the probabilities given by its centroid are similar to the second-order probabilities of all the in-links assigned to that cluster. That is, there does not exist an in-link such that the difference between a second-order probability and the corresponding centroid probability is greater than the specified threshold, $\gamma$ (see Definition 3);

   (f) If the solution is not accurate, run $K$-means again, now with $K := K + 1$ (for $K \leq I$ since for $K = I$ the solution is necessarily exact). In practice we make $K$ vary from $K = 2$ to $K = I$ according to a geometric progression ($K := K^2$), otherwise for a state with a large number of in-links, the number of iterations of the clustering algorithm would lead to a prohibitive running time;

3. Create $K - 1$ clones of state $A_x$, distribute the in-links among the clones according to the clustering assignment. Compute the weights of the out-links according to Definition 5.

Figure 6: The clustering-based cloning method for a state $A_x$.

is assigned to the cluster corresponding to the clone and the three other links assigned to the cluster corresponding to the original state. Since link $(A_2, A_5)$ is closer to the first cluster it is assigned to it and both centroids are recomputed. Figure 8 (c) represents the solution where the link $(A_2, A_5)$ is reassigned and the $K$-means algorithm reaches a solution. The comparison of the first-order probabilities given by the centroids, and the second-order probabilities indicated in Figure 8 (a), shows that the solution obtained presents a nice tradeoff between number of states and the model's accuracy.

When a state, $A_x$, is being evaluated, if a state that links to it was previously cloned, state $A_x$ may have in-links from both a state and its clones. Links from a state and its clones possess the same second-order probabilities with respect to the in-links to state $A_x$. Therefore, step 2 (b) in Figure 6 will place such in-links in the same cluster due to the condition of placing identical vectors in the same cluster. Since the vectors are identical the in-links will move among clusters as a unit. For example, state $A_7$ in Figure 8 (c) has in-links from $A_5$ and $A_5'$. Transitions $(A_5, A_7)$ and $(A_5', A_7)$ induce the same second-order probabilities for transitions to state $A_7$. As a result, the solution obtained for state $A_7$ would be the same if that state was evaluated before state $A_5$. Therefore, we can say that the second-order model obtained by the proposed method is independent of the order on which the states are evaluated.

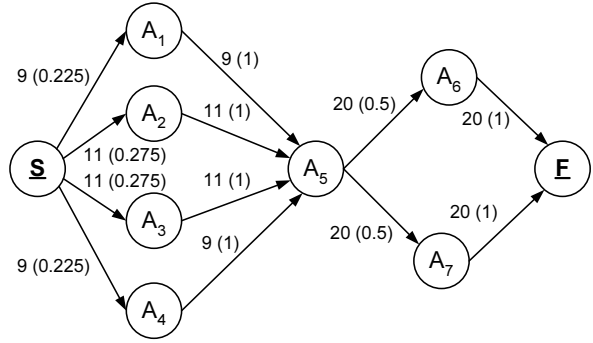

| Session | NOS | $p_{i,j\,k}$ |
|---|---|---|
| $A_1, A_5, A_6$ | 6 | $p_{1,5\,6} = 6/9$ |
| $A_1, A_5, A_7$ | 3 | $p_{1,5\,7} = 3/9$ |
| $A_2, A_5, A_6$ | 7 | $p_{2,5\,6} = 7/11$ |
| $A_2, A_5, A_7$ | 4 | $p_{2,5\,7} = 4/11$ |
| $A_3, A_5, A_6$ | 4 | $p_{3,5\,6} = 4/11$ |
| $A_3, A_5, A_7$ | 7 | $p_{3,5\,7} = 7/11$ |
| $A_4, A_5, A_6$ | 3 | $p_{4,5\,6} = 3/9$ |
| $A_4, A_5, A_7$ | 6 | $p_{4,5\,7} = 6/9$ |

Figure 7: An example to illustrate the clustering-based cloning method.

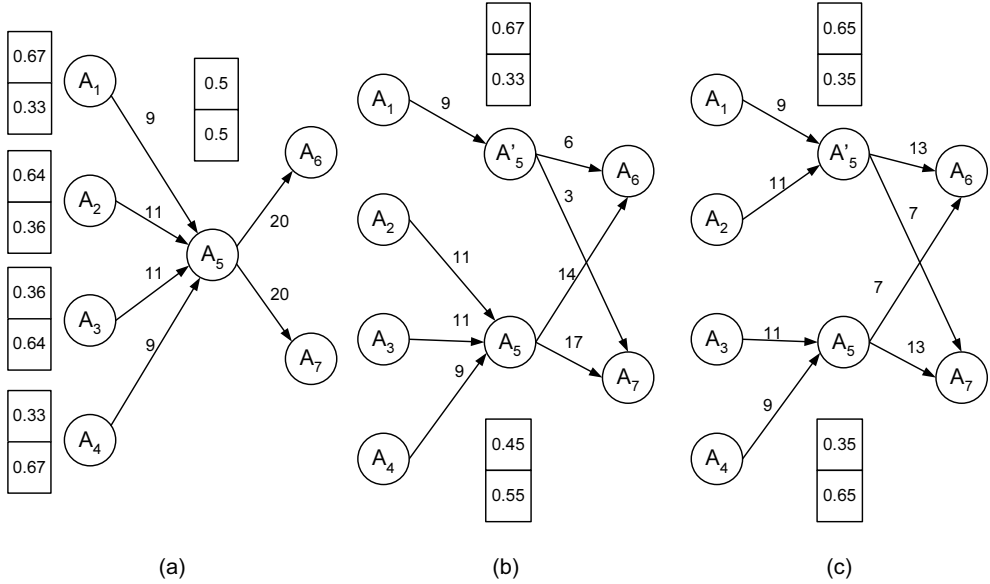

Figure 8: Clustering-based cloning of state $A_5$ in the example given in Figure 7.

We note that, by setting the accuracy threshold parameter, $\gamma$, the analyst is able to control the model complexity, measured by the number of states, while setting the intended accuracy. If $\gamma = 0$ all the resulting second-order probabilities will be exact; if $\gamma = 1$ no cloning is performed and the resulting model will correspond to a first-order model; for other values of $\gamma$ a model between the two scenarios described is obtained.

We also note that a model representing second-order probabilities implicitly represents the first-order probabilities. For example, in Figure 8 (a) we have $p_{5,6} = 0.5$ that is equivalent to $\frac{9}{40}0.67 + \frac{11}{40}0.64 + \frac{11}{40}0.36 + \frac{9}{40}0.33 = 0.5$ and corresponds to the weighted average for the second-order probabilities.

## 4 Experimental Evaluation

In order to assess the properties of the proposed method we conducted a set of experiments on both real and artificial data. Experiments with real data allow us to evaluate the model in a real-world scenario, while experiments with artificially generated data allow us to evaluate the model in a wider variety of possible scenarios. We will now describe our method to generate artificial data sets.

## 4.1 A Method to Generate Artificial Log Files

The method for generating artificial log files is divided into two stages: (i) creating a web topology, and (ii) creating a collection of user sessions based on the generated topology.

It has been found that the distribution of the number of out-links from a web page can be explained by a power-law with exponent 2.72, and that the number of in-links to a web page can be explained by a power-law with exponent 2.1; see [Adamic and Huberman, 2002]. Thus, in our method to generate a web topology that mirrors a web site there are three parameters: the number of pages in the site, $N$, the exponent for the power-law describing the in-link distribution and the exponent for the power-law describing the out-link distribution.

The method works as follows. For each of the $N$ pages we generate its number of out-links and in-links according to the corresponding power-law distributions. At this stage, the in-links generated for a given state are called *unmatched in-links* and the out-links are called *unmatched out-links*. From the collection of all unmatched in-links we pick at random one in-link and from the collection of unmatched out-links we pick at random one out-link. The matching of the in-link to the out-link forms a candidate link, which is promoted to a link if it is a *valid link*. A candidate link is considered to be valid if it is not a loop and does not yet exist a link with the same source state and target state.

After creating a web topology it is necessary to generate a collection of user navigation sessions. In order to simulate a user's choice of link to follow next, we compute the PageRank [Brin and Page, 1998] for each page on the generated topology, as a measure of the user's interest in each web page. Then, we compute the vector of initial probabilities and the transition probabilities with respect to the topology. The initial probability of a page is estimated by its PageRank divided by the sum of the PageRanks of all pages, and transition probabilities are estimated according to the PageRank random surfer model.

In order to generate a collection of sessions, we set the intended number of sessions via a parameter. As suggested in [Huberman et al., 1998] we employ a power-law distribution with exponent 1.5 to set the number of user clicks, $L$, in a given session. For each session, we randomly choose the start state according to the distribution of the the initial probabilities and each of the $L$ links is chosen according to the probabilities of the out-links of the current state. Also, as in the random surfer model, at each state there is a probability of 0.15 for the session terminating.

## 4.2 Experiments with Random Data

We have conducted a set of experiments with random data to assess the model's behaviour. The number of states parameter for the web topology generation was set with the values $\{1000, 3000, 6000, 9000, 12000, 15000, 20000\}$. The number of sessions generated for a given topology was set to be twice the number of states in the topology, and the accuracy threshold parameter, $\gamma$ (see Definition 3), was set with the values $\{0, 0.1, 0.2, 0.4, 0.6, 0.8, 1\}$. Each configuration was repeated 10 times so that every result presented corresponds to the average of ten runs. We also provide a comparison with the corresponding $N$-gram model for $N$=3, 4 and 5. Table 2 presents a summary of the characteristics for the random data sets generated. The column headers represent the number of states given as an input parameter for the topology generator, and the row entry corresponding to the number of states gives the number of states covered by the navigation sessions generated.

|  | 1000 | 3000 | 6000 | 9000 | 12000 | 15000 | 20000 |
|---|---|---|---|---|---|---|---|
| Num. states | 989 | 2961 | 5916 | 8862 | 11814 | 14765 | 19703 |
| Num. sessions | 13002 | 38329 | 74489 | 112210 | 151069 | 188233 | 250130 |
| Num. requests | 75488 | 222394 | 427417 | 646012 | 869676 | 1085314 | 1442793 |
| Avg. session length | 5.8 | 5.8 | 5.7 | 5.8 | 5.8 | 5.8 | 5.8 |
| Stdev. session length | 29.8 | 30.0 | 29.3 | 29.5 | 29.5 | 29.6 | 29.7 |
| Max. session length | 62.6 | 63.9 | 72.2 | 68.8 | 72.7 | 76.8 | 75.3 |
| Num. starting states | 967 | 2895 | 5777 | 8656. | 11529 | 14411 | 19244 |
| Num. terminating states | 782 | 2252 | 4423 | 6564 | 8627 | 10935 | 14688 |
| Avg. num. out-links/page | 3.4 | 3.3 | 3.3 | 3.3 | 3.2 | 3.3 | 3.3 |
| Stdev. num. out-links/page | 1.9 | 2.0 | 2.0 | 1.9 | 2.0 | 1.9 | 2.2 |
| Avg. num. in-links/page | 3.6 | 3.6 | 3.5 | 3.5 | 3.5 | 3.5 | 3.5 |
| Stdev. num. in-links/page | 8.6 | 11.4 | 11.1 | 11.3 | 11.9 | 11.6 | 11.4 |

Table 2: Summary statistics for the random data sets.

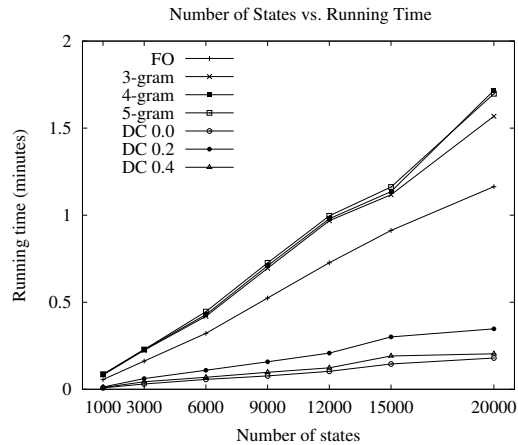

Figure 9: The running time with the number of states in the web topology for building several variations of the HPG model.

Figure 9 shows the variation of the running time for building the model with a given number of states in the web topology. The results confirm that both the first-order model and the $N$-gram model are approximately linear with the number of states and that the additional time necessary for the dynamic clustering-based method is relatively short. Note that for the dynamic clustering-based method (DC in the figures) we indicate the running time for three different values of the accuracy threshold parameter, and the running time indicated is the additional time above that spent in constructing the first-order model. The method is fastest for $\gamma = 0$ because in that case the clustering step is not required. In fact, $\gamma = 0$ implies that only probabilities that are the same are clustered together, and since we force the initial solution to place identical vectors of probabilities in the same cluster, the initial solution corresponds to the final solution if $K$ is set to be the number of distinct probability vectors (where $K$ is the number of clusters).

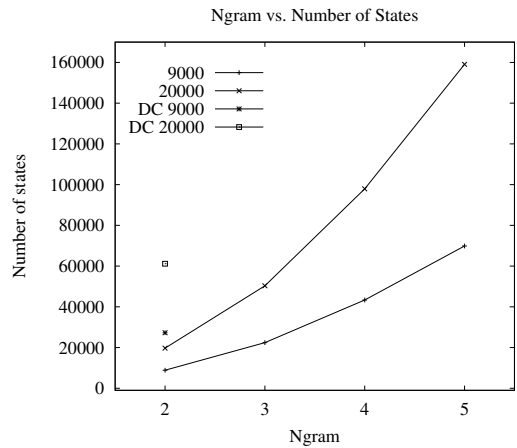

Figure 10: The number of states for different values of $N$ in the $N$-gram model and for the dynamic clustering-based method for $\gamma = 0$.

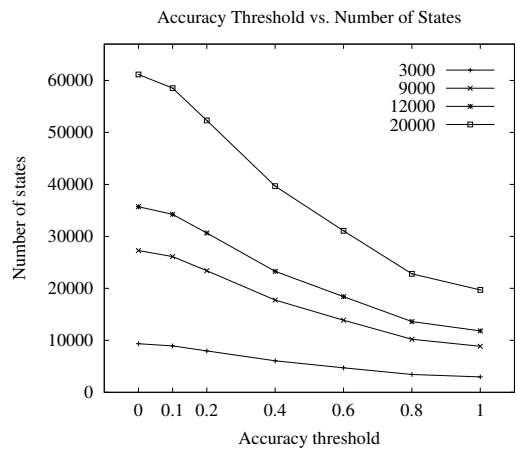

Figure 11: The number of states for different values of the accuracy threshold parameter, $\gamma$, on the dynamic clustering-based method.

Figure 10 shows the variation of the number of states in the model for different configurations of the $N$-gram model and compares it with the corresponding dynamic clustering-based method when the accuracy threshold parameter is set to $\gamma = 0$, that is, requiring full precision for second-order probabilities. The results show that the dynamic clustering-based method induces more states than the 3-gram model but significantly less than higher order $N$-grams. Note that $N = 2$ corresponds to the first-order model, and that the number of states corresponding to the dynamic clustering-based method are represented by a single point, since the variation of the $N$ parameter does not influence the results in this case.

Figure 11 shows the variation of the number of states induced by the dynamic clustering-based method for different values of the accuracy threshold parameter, $\gamma$. Note that $\gamma = 0$ corresponds to maximum precision and $\gamma = 1$ to maximum flexibility; the latter case corresponds to the first-order model. The results show that it is possible to have some control over the number of states induced by setting the value of the accuracy threshold parameter, since the variation is close to being linear.

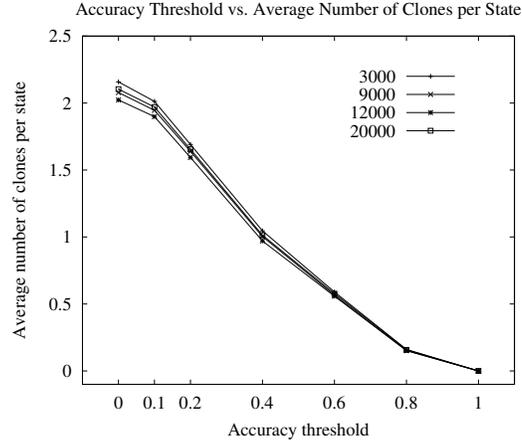

Figure 12: The average number of clones per state for different values of the accuracy threshold parameter, $\gamma$, on the dynamic clustering-based method.

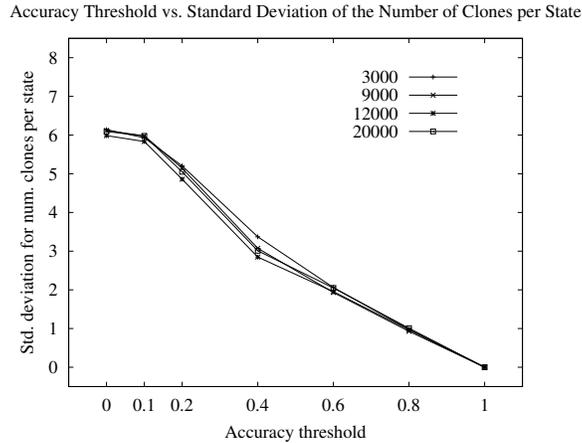

Figure 13: The standard deviation for the number of clones per state for different values of the accuracy threshold parameter, $\gamma$, on the dynamic clustering-based method.

Figure 12 shows the average number of clones per state and Figure 13 shows the standard deviation of the number of clones per state. It can be seen that the average and standard deviation are insensitive to the model's number of states and depends only on the accuracy threshold. Moreover, despite each state having only 2 clones on average, the standard deviation is high when $\gamma$ is low.

## 4.3 Experiments with Real Data

In our experiments we use two different data sets. The first is from a university site and was made available by the authors of [Berent et al., 2001]. This data is based on a random sample of two weeks of usage data from a university site during April 2002. According to the authors, during the time of collection caching was prohibited by the site and the site was cookie based in order to ease user identification. The data was made available with the sessions already identified; according to [Berent et al., 2001] sessions were inferred based on the cookie information. The second data set was obtained from the authors of [Perkowitz and Etzioni, 2000] and corresponds to ten days of usage records from the site http://machines.hyperreal.org/ in 1999. The data made available was organized in sessions and, according to the authors, caching was disabled during collection. Table 3 presents a summary of the characteristics of the two log files.

| Characteristic | University | Hyperreal |
|---|---|---|
| Num. states | 9149 | 7319 |
| Num. sessions | 20950 | 14227 |
| Num. requests | 106991 | 104551 |
| Avg. session length | 5.1 | 7.3 |
| Stdev. session length | 92.1 | 743.8 |
| Max. session length | 396 | 2179 |
| Num. starting states | 479 | 1376 |
| Num. terminating states | 1975 | 1887 |
| Avg. num. out-links/page | 3.1 | 4.2 |
| Stdev. num. out-links/page | 13.3 | 15.3 |
| Avg. num. in-links/page | 2.9 | 4.1 |
| Stdev. num. in-links/page | 16.0 | 15.7 |

Table 3: Summary statistics for the real data set.

| | | FO (2-gram) | 3-gram | 4-gram | 5-gram | DC, $\gamma = 0$ |
|---|---|---|---|---|---|---|
| University | Time (min.) | 0.28 | 0.38 | 0.30 | 0.41 | 0.56 |
| | Num. states | 9149 | 26644 | 41643 | 48424 | 21837 |
| Hyperreal | Time (min.) | 0.25 | 0.32 | 0.26 | 0.34 | 1.34 |
| | Num. states | 7319 | 28666 | 46637 | 56589 | 25839 |

Table 4: Running time relative to the first-order model and number of states for the first-order model, three $N$-gram models and the dynamic clustering-based model with $\gamma = 0$.

Table 4 shows the number of states and the running time (in minutes) for the first-order model, three $N$-gram configurations and the dynamic clustering-based method with $\gamma = 0$. The results show that the number of states induced by the dynamic clustering-based method is smaller than any of the $N$-gram models but with the cost of a higher running time.

Tables 5 and 6 show the effect that the accuracy parameter, $\gamma$, has on the running time and on the number of clones induced. As was already verified with random data, the dynamic clustering-based method is fastest for $\gamma = 0$ when the clustering step is not required. The number of cluster $K$ for the $K$-means algorithm was set to vary as $K := 2K$ and $K := K^2$.

The results suggest that, especially for the hyperreal data set, the gain in varying $K$ smoothly (measured by a smaller number of states due to a finer search) is not evident when compared with the decrease in performance. For real data, the total number of clones and the average number of clones per state are much more stable than for random data. The reason for this could be that, for real data, different sessions tend to be disjoint, that is, sessions go through the same states but follow different in- and out-links. In such a scenario, second-order probabilities diverge significantly from the corresponding first-order probabilities and cloning is required even if the accuracy threshold parameter is set to a high value.

| | | University Data | | | | | |
|---|---|---|---|---|---|---|---|
| | | $\gamma = 0$ | $\gamma = 0.1$ | $\gamma = 0.2$ | $\gamma = 0.4$ | $\gamma = 0.6$ | $\gamma = 0.8$ |
| $K := K^2$ | Time (min.) | 0.56 | 1.18 | 1.20 | 1.17 | 1.08 | 1.1 |
| | Num. states | 21837 | 21831 | 21800 | 21619 | 20355 | 18460 |
| | Avg. clones/state | 1.39 | 1.39 | 1.38 | 1.36 | 1.22 | 1.02 |
| | Stdev. clones/state | 7.51 | 7.51 | 7.51 | 7.50 | 7.49 | 7.44 |
| $K := 2K$ | Time (min.) | 0.56 | 1.78 | 1.81 | 1.83 | 1.89 | 1.91 |
| | Num. states | 21837 | 21828 | 21796 | 21601 | 20338 | 18305 |
| | Avg. clones/state | 1.39 | 1.39 | 1.38 | 1.36 | 1.22 | 1.00 |
| | Stdev. clones/state | 7.51 | 7.51 | 7.51 | 7.50 | 7.49 | 7.41 |

Table 5: Running time, total number of states and average number of clones per state for several values of the accuracy threshold parameter, $\gamma$, for the university data.

| | | Hyperreal Data | | | | | |
|---|---|---|---|---|---|---|---|
| | | $\gamma = 0$ | $\gamma = 0.1$ | $\gamma = 0.2$ | $\gamma = 0.4$ | $\gamma = 0.6$ | $\gamma = 0.8$ |
| $K = K^2$ | Time (min.) | 1.34 | 2.62 | 2.64 | 2.66 | 2.67 | 3.03 |
| | Num. states | 25839 | 25828 | 25806 | 25664 | 24140 | 21046 |
| | Avg. clones/state | 2.53 | 2.53 | 2.53 | 2.50 | 2.30 | 1.88 |
| | Stdev. clones/state | 10.72 | 10.72 | 10.72 | 10.71 | 10.73 | 10.69 |
| $K := 2K$ | Time (min.) | 1.34 | 5.24 | 5.40 | 5.40 | 5.45 | 5.49 |
| | Num. states | 25839 | 25825 | 25801 | 25596 | 24035 | 20669 |
| | Avg. clones/state | 2.53 | 2.53 | 2.53 | 2.50 | 2.28 | 1.82 |
| | Stdev. clones/state | 10.72 | 10.71 | 10.71 | 10.71 | 10.72 | 10.63 |

Table 6: Running time, total number of states and average number of clones per state for several values of the accuracy threshold parameter, $\gamma$, for the hyperreal data.

The results obtained for real data differ from those obtained for random data in revealing higher running times for the dynamic clustering-based method than those indicated by the simulations on random data. This may be explained by the fact that on real data the navigation does not precisely follow the PageRank random surfer model, which is inherently first-order. When the behaviour is first-order the choice of the next link does not depend on which in-link was followed to get to the state. Therefore, by the law of large numbers, in a first-order model the in-links to a state tend to have identical second-order probabilities, resulting in a smaller number of clusters being necessary to accurately model any second-order probabilities.

We also note the higher variability shown in the real data set. The statistics given in Tables 2 and 3 show that the real data has a much higher variance for session length and for the number of out-links per state than the random data. The number of clones per state affects the running time due to method used to find the number of clusters $K$ needed to reach a certain accuracy. In Table 4 it can be seen that the clustering-based method penalises the hyperreal data, and this fact is probably related to the higher variability of the number of clones per state, as can be seen in Tables 5 and 6. In addition, for the clustering-based method with $\gamma = 0$, the maximum number of clones per state is 242 for the random data with 9000 states, is 331 for the university data and 587 for the hyperreal data.

We further note that we should not directly compare the running time of $N$-gram models with that of the dynamic clustering-based model. It is possible to take, for example, a 2-gram, a 3-gram and a 4-gram model and weight them in an appropriate manner in order to induce the probability of a given trail. However such weights would have the same value for every state in the model, and therefore, every state would be dealt with in the same way when determining its required history depth. On the other hand, the dynamic clustering-based model assesses the required history depth for each state individually, therefore being a more demanding method.

## 5 Evaluation

In this section we provide a preliminary evaluation of the dynamic clustering-based method. First, we discuss the method's performance. The running times indicated in Section 4 correspond to the additional time required, above the time necessary to build the first-order model. The results shown in Figure 9 obtained from experiments with random data suggest that our method is close to linear in the size of the model and that the sum of the times for running the first-order and dynamic clustering-based methods is comparable to that of the $N$-gram model.

The results with real data were not consistent with those obtained with random data with respect to the comparison to the $N$-gram model. On the one hand, with random data the running time for the dynamic clustering-based method was comparable to the running time required by the $N$-gram model, while with real data the $N$-gram is faster. On the other hand, with random data the number of states induced by the new method was higher than for the 3-gram model but smaller that the 4-gram model, and for real data the number of states induced is smaller than the 3-gram model.

We believe that the disparity between the experiments with random and real data is due to the fact that the random data was generated according to the random surfer model (which is inherently first-order) and with such model link following does not take second-order probabilities into account. If every in-link to a state holds a similar probability distribution with respect to the choice of the the next link to follow, the dynamic clustering-based algorithm will tend to converge at a faster rate since the number of clusters, $K$, necessary to accurately represent second-order probabilities tends to be smaller.

Table 7 reports the number of sessions dropped by each of the $N$-gram models we have considered. To obtain the last row in the table we assume that session length follows a power-law distribution with exponent 1.5 [Huberman et al., 1998], and thus the probability that a session be of length $L$ is $L^{(-1.5)}/zeta(1.5)$, where the $zeta$ function provides the normalization constant for the probabilities. As was mentioned in Section 2.2, when using the 3-gram model sessions of length one are dropped. According to the above estimate, on average, over 38.3% of the sessions are expected to be first-order but not second-order and are, therefore, dropped from the model. For the real data used in our experiments the percentage of first-order sessions dropped was smaller than expected, but more than 29% in both cases, see Table 7. On the other hand, the dynamic clustering-based method does not drop any session, as it takes full advantage of longer sessions if they are available and uses all existing short sessions.

| Dataset | Sessions Dropped | FO (2-gram) | 3-gram | 4-gram | 4-gram |
|---|---|---|---|---|---|
| University | Number of Sessions | 0 | 7075 | 9763 | 11576 |
| | % of Sessions | 0.0% | 33.77% | 46.60% | 55.26% |
| Hyperreal | Number of Sessions | 0 | 4153 | 6386 | 7765 |
| | % of Sessions | 0.0% | 29.19% | 44.89% | 54.58% |
| Theoretical | % of Sessions | 0.0% | 38.27% | 51.80% | 59.17% |

Table 7: Amount of sessions dropped by the $N$-gram model.

Finally, we discuss the quality of the rules induced by a HPG created by the dynamic clustering-based method. The set of rules with probability above a specified parameter, called its cut-point, can be induced by an algorithm such as the one given in [Borges and Levene, 2004]. The rules returned when using dynamic clustering will be a proper subset of the ones returned for the first-order model, with the sequences of pages returned, called trails, having higher probability than the ones returned by the first-order model. This statement can be illustrated by comparing the models given in Figures 1 and 3. After cloning is applied there are less trails available and each having higher probability than the corresponding trail in the first-order model. For example, in Figure 1 the trail $A_1A_2A_6$ (which was not traversed by any user) has probability 0.0625 but the same trail is not included in the model given in Figure 3. Also, the trail $A_1A_2A_3$ (which was traversed three times by a user) has probability 0.1875 in Figure 3 but has probability 0.375 in Figure 3.

## 6  Concluding Remarks

We have presented a method to extend a first-order Markov model such that it incorporates second-order probabilities. The method makes use of the state cloning concept where a state is duplicated if the first-order probabilities induced by its out-links diverge significantly from the corresponding second-order probabilities. The method proposes the use of a clustering algorithm to identify the best way to distribute a state's in-links between the state and its clones. Results of experiments conducted on both real and synthetic data were presented suggesting that the method's performance is linear time in the size of the model and that the increase in the number of states is close to that induced by the 2-gram model but smaller than for higher order $N$-gram models.

We are planning to provide a more complete evaluation of the benefits of the technique in terms of increased modelling accuracy, and are working on the generalization of the method for higher orders.